\input{aipcheck}
\documentclass[final]{aipproc}
\layoutstyle{6x9}
\usepackage{amssymb}

\newcommand{\ftn}{\footnotesize}

\newcommand{\TeV}{{\mbox{\rm TeV}}}
\newcommand{\MeV}{{\mbox{\rm MeV}}}
\newcommand{\GeV}{{\mbox{\rm GeV}}}

\newcommand{\vrho}{{\mbox{$\bar\rho$}}}

\def\openep{\leavevmode\hbox{\normalsize$\iota$\kern-3.8pt$^$-}}
\def\vtau{\leavevmode\hbox{\normalsize$\tau$\kern-5.pt$\iota$}}
\def\vtauf{\leavevmode\hbox{\ftn$\tau$\kern-4.pt$\iota$}}

\def\beq{\begin{equation}}
\def\eeq{\end{equation}}
\def\bea{\begin{eqnarray}}
\def\eea{\end{eqnarray}}

\newcommand{\Ti}{\ensuremath{T_{\rm I}}}
\newcommand{\Tkr}{\ensuremath{T_{\rm KR}}}
\newcommand{\Ts}{\ensuremath{T_{\rm SUSY}}}
\newcommand{\Tns}{\ensuremath{T_{\rm NS}}}
\newcommand{\Tc}{\ensuremath{T_{\rm C}}}

\newcommand{\Omax}{{\mbox{$\Omega_{\ax} h^2$}}}
\newcommand{\Omgr}{{\mbox{$\Omega_{\Gr} h^2$}}}
\newcommand{\Omx}{{\mbox{$\Omega_\chi h^2$}}}
\newcommand{\Omqns}{\ensuremath{\Omega_q^{\rm NS}}}

\newcommand\Gm[1]{\Gamma_{#1}}

\newcommand{\gl}{\ensuremath{\tilde{g}}}
\newcommand{\ax}{\ensuremath{\tilde{a}}}
\newcommand{\sq}{\ensuremath{\tilde{q}}}
\newcommand{\Gr}{\ensuremath{\tilde{G}}}
\newcommand{\nequ}{\ensuremath{n^{\rm eq}}}

\newcommand{\sFref}[2]{Fig.~\ref{#1}-${\sf ({#2})}$}

\newcommand{\Eref}[1]{Eq.~(\ref{#1})}

\begin{document}

\title{Quintessential Kination
and Thermal Production of SUSY \emph{e}-WIMPs}

\classification{98.80.Cq,98.80.-k, 95.35.+d}

\keywords{Cosmology, Dark Energy, Dark Matter}

\author{M.E. G\'omez}{
  address={
Departamento de Fisica Aplicada, Universidad de Huelva,  21071
Huelva, Spain
    }
}

\author{S. Lola}{
  address={
Department of Physics, University of Patras, 26500 Patras, Greece
}}

\author{C. Pallis}{
  address={
Departamento de Fisica Aplicada, Universidad de Huelva,  21071
Huelva, Spain
    }
}

\author{J. Rodr\'iguez-Quintero}{
  address={
Departamento de Fisica Aplicada, Universidad de Huelva,  21071
Huelva, Spain
    }
}

\begin{abstract}
The impact of a \emph{kination-dominated} (KD) phase generated by
a quintessential exponential model on the thermal abundance of
\emph{Supersymmetric} (SUSY) \emph{extremely Weekly Interacting
Massive Particles} (\emph{e}-WIMPs) is investigated. For values of
the quintessential energy-density parameter at the eve of
nucleosynthesis close to its upper bound, we find that (i) the
gravitino ($\Gr$) constraint is totally evaded for unstable
$\Gr$'s; (ii) the thermal abundance of stable $\Gr$ is not
sufficient to account for the \emph{cold dark matter} (CDM) of the
universe; (iii) the thermal abundance of axinos ($\ax$) can
satisfy the CDM constraint for values of the initial
(``reheating'') temperature well above those required in the
\emph{standard cosmology} (SC).

\end{abstract}

\maketitle

\section{Introduction}\label{intro}

A plethora of data \cite{wmap} indicates that the two major
components of the universe are CDM and \emph{Dark Energy} (DE).
The DE component can be explained with the introduction of a
slowly evolving scalar field, $q$, called quintessence
\cite{early}. An open possibility in this scenario is the
existence of an early KD era \cite{kination}, where the universe
is dominated by the kinetic energy of $q$. During this era, the
expansion rate of the universe increases \emph{with respect to}
(w.r.t.) its value in SC. As a consequence, the relic abundance of
WIMPs (e.g., the \emph{lightest neutralino}) is also significantly
enhanced \cite{salati, jcapa}.

WIMPs are the most natural candidates for the second  major
component of the universe, CDM. In addition, supersymmetric
theories predict the existence of even more weakly interacting
massive particles, known as \emph{e}-WIMPs \cite{ewimps}. These
are the gravitino and the axino (SUSY partners of the graviton and
the axion respectively). Their interaction rates are suppressed by
the reduced Planck scale, $m_{\rm P}=M_{\rm P}/\sqrt{8\pi}$ (where
$M_{\rm P}=1.22\times10^{19}~{\rm GeV}$ is the Planck mass) in the
case of $\Gr$ and by the axion decay constant,
$f_a\sim(10^{10}-10^{12})~{\rm GeV}$ in the case of $\tilde a$.
Due to the weakness of their interactions, e-WIMPs depart from
chemical equilibrium very early and their relic density is diluted
by  primordial inflation. However, they can be reproduced in two
ways: (i) in the thermal bath, through scatterings and decays
involving superpartners \cite{prod1, prod0, prod2}, and (ii)
non-thermally from the decay of the next-to-lightest
supersymmetric particle; in this case the results are highly model
dependent.

In this talk, which is based on Ref.~\cite{GLPR}, we reconsider
the creation of a \emph{quintessential kination scenario} (QKS) in
the context of the exponential quintessential model \cite{jcapa,
brazil}, and investigate its impact on the thermal production of
\emph{e}-WIMPs.

\section{The Quintessential Exponential Model} \label{quint}

The quintessence field, $q$, of our model satisfies the equation:
\beq \ddot q+3H\dot q+dV/dq=0,~~\mbox{where}~~V=V_0 e^{-\lambda
q/m_{_{\rm P}}}\label{qeq} \eeq
is the adopted potential, dot denotes derivative w.r.t the cosmic
time $t$ and $H$ is the Hubble parameter, $H\simeq\sqrt{\rho_q
+\rho_{_{\rm R}}}/\sqrt{3}m_{_{\rm P}}$ where $\rho_q=\dot
q^2/2+V$ and $\rho_{_{\rm R}}=\pi^2g_*T^4/30$ the radiation energy
density with $g_*$ the effective number of massless degrees of
freedom.

We impose on our quintessential model the following constraints:
\begin{enumerate}
\item \label{domk} \emph{Initial Domination of Kination.} We focus
our attention in the range of parameters with $\Omega^{\rm
I}_q=\Omega_q(\Ti)\gtrsim0.5$ where
$\Omega_q\simeq\rho_q/(\rho_q+\rho_{{\rm R}})$ is the
quintessential energy-density parameter.

\item\label{nuc} \emph{Nucleosynthesis (NS) Constraint.} At the
onset of NS, $\Tns=1~\MeV$, $\rho_q$ is to be sufficiently
suppressed w.r.t $\rho_{_{\rm R}}$, i.e., \cite{oliven}
$\Omega_q^{\rm NS}=\Omega_q(\Tns)\leq0.21$ at $95\%$ c.l.

\item\label{para} \emph{Inflationary Constraint.} Assuming that
the power spectrum of the curvature perturbations is generated by
an early inflationary scale, an upper bound on the initial value
of $H$, $H_{\rm I}$, can be obtained, namely $H_{\rm
I}\lesssim2.65\times10^{14}~\GeV$.

\item\label{rhoq0} \emph{Cosmic Coincidence Constraint.} The
present value of $\rho_q$, $\rho^0_q$, must be compatible with the
preferred range for DE, implying $\Omega^0_q = 0.74.$

\item\label{wq} \emph{Acceleration Constraint.} Successful
quintessence has to account for the present-day acceleration of
the universe, i.e. $-1\leq w_q(0)\leq-0.86~~\mbox{($95\%$ c.l.)}$,
where $w_q=(\dot q^2/2-V)/(\dot q^2/2+V)$ is the barotropic index
of the $q$-field.

\end{enumerate}

\begin{figure}[!t]\centering\hspace*{0.2cm}
\includegraphics[width=6.cm,angle=-90]{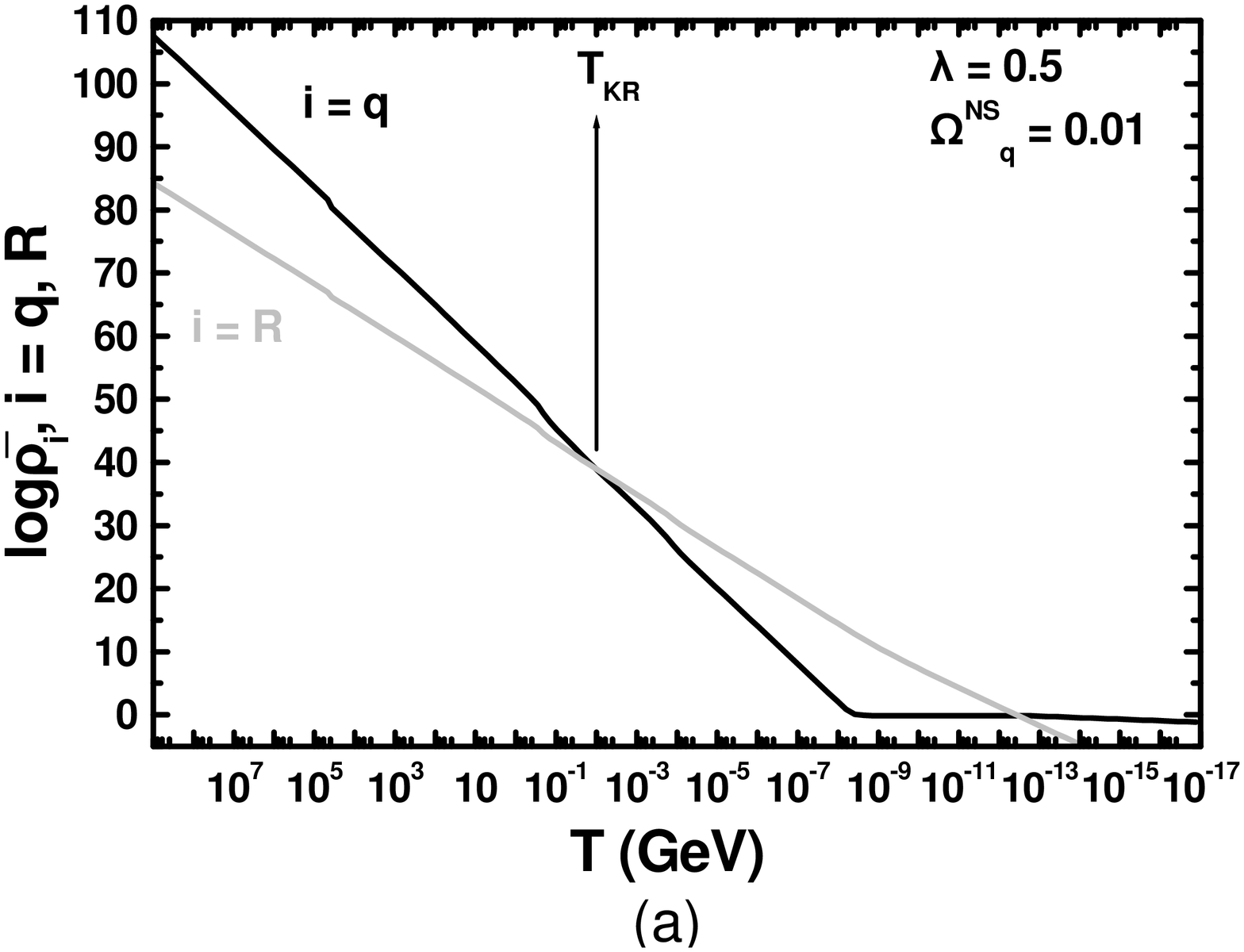}\hspace*{-1cm}
\includegraphics[width=6.cm,angle=-90]{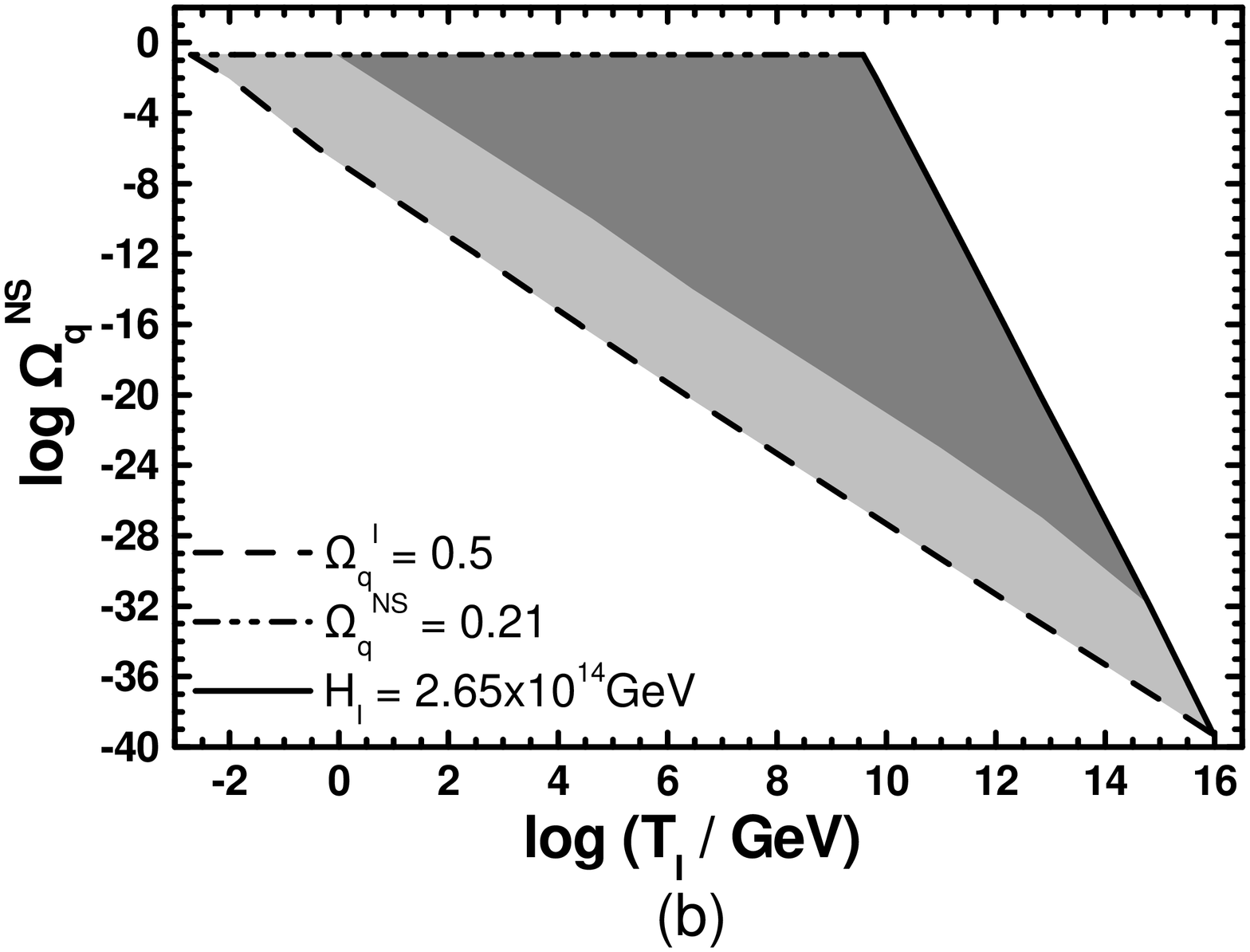}
\caption{\sl (a) The evolution of $\log\vrho_i$
($\vrho_i=\rho_{i}/\rho^0_{\rm c}$ with $\rho^0_{\rm
c}=8.099\times10^{-47}h^2~{\rm GeV^4}$ and $h=0.72$) with $i=q$
(black line) and R (light gray line) as a function of $T$ for
$\lambda=0.5$, $\Ti=10^9~\GeV$ and $\Omqns=0.01$. (b) Allowed
(gray and lightly gray shaded) region by the conditions \ref{domk}
-- \ref{para} in the $\log\Ti-\log\Omqns$ plane.} \label{r5}
\end{figure}

Solving \Eref{qeq} with $q(\Ti)=0$ and $\dot q(\Ti)$ such that the
condition \ref{domk} is satisfied, we find that, during its
evolution, $q$ undergoes three phases (see \sFref{r5}{a}, where we
plot $\log\vrho_i$ with $i=q$ and R versus $T$ for
$\Ti=10^9~\GeV$, $\Omqns=0.01$ and $\lambda=0.5$):

\begin{itemize}

\item The {\em KD phase} \cite{kination}, where $\rho_q\simeq \dot
q^2/2$, implying $w_q\simeq1$ and thus, $\rho_q\propto T^6$. The
transition from kination to radiation occurs at $\Tkr$ such that
$\rho_q(\Tkr)=\rho_{\rm R}(\Tkr)$ -- see \sFref{r5}{a}. Solving
this we find $\Tkr=\Tns\left({g^{\rm NS}_{*}/g^{\rm
KR}_{*}}\right)^{1/2}\left({(1-\Omega^{\rm NS}_q)/\Omega^{\rm
NS}_q}\right)^{1/2}$.

\item The {\em frozen-field dominated phase}, where the universe
becomes radiation dominated and $\rho_q$ is dominated initially by
$\dot q/2$ and then by $V$.

\item The {\em late-time attractor dominated phase} during which $\rho_q\simeq V$
dominates the universal evolution, with $w_q\simeq w^{\rm
fp}_q=\lambda^2/3-1$ for $\lambda<\sqrt{3}$.

\end{itemize}
Today we obtain a transition from the frozen-field to the
attractor dominated phase. Although this does not provide a
satisfactory resolution of the coincidence problem, the
observational data can be reproduced. In particular, satisfying
the condition \ref{wq} implies $\lambda\leq0.9$ whereas the
condition \ref{rhoq0} can be fulfilled by conveniently adjusting
$V_0$ \cite{jcapa, brazil}. These two constraints are independent
of the parameters $\Ti$ and $\Omqns$, which can be restricted by
the requirements \ref{domk} -- \ref{para}. In \sFref{r5}{b}, we
present the allowed region of our model in the
$\log\Ti-\log\Omqns$ plane (shaded in gray and light gray). We
observe that for a reasonable set of the parameters ($\lambda,
\Ti, \Omqns$), our model can become consistent \cite{jcapa,
brazil} with the observational data.

%%%%%%%%%%%%%%%%%%%%%%%%%%%%%%%%%%%%%%%%%%%%%%%%%%%%%%%%%%%%%%%%%%%%
%%%%%%%%%%%%%%%%%%%%%%%%%%%%%%%%%%%%%%%%%%%%%%%%%%%%%%%%%%%%%%%%%%%%

\section{Thermal production of SUSY $e$-WIMP${\rm s}$}

The number density $n_{\chi}$ of a SUSY \emph{e}-WIMP $\chi$
(where $\chi$ stands for $\Gr$ or $\ax$) satisfies the Boltzmann
equation \cite{prod0, prod2, axino} which can be written as
\beq \dot n_\chi+3Hn_\chi=C_\chi n^{{\rm eq}2}+
\sum_i{g_i\over2\pi^2}m_i^2\,T\,K_1(m_i/T)\,\Gm{i}. \label{nx}\eeq
Here $\nequ={\zeta(3)T^3/\pi^2}$ is the equilibrium number density
of the bosonic relativistic species, $m_i$ [$g_i$] is the mass
[number of degrees of freedom] of the particle $i$ and $K_n$ is
the modified Bessel function of the 2nd kind. In the relativistic
regime ($T\gg m_i$) $C_\chi$ has been calculated \cite{prod1,
prod2} using the \emph{Hard Thermal Loop Approximation}, resulting
to $C_\chi=C_\chi^{\rm HT}$, where
\beq C_\chi^{\rm HT} =\left\{\matrix{
\left(3\pi/16\zeta(3)m^2_{\rm P}\right)\sum_{\alpha=1}^{3}
\left(1+{M_\alpha^2/3 m_{\chi}^2}\right)c_\alpha g_\alpha^2
        \ln\left({k_{\alpha}/g_\alpha}\right)
\hfill \mbox{for}  & \chi=\tilde G, \hfill \cr
\left(27g_3^4/\pi^3f^2_a\zeta(3)\right) \ln\left(1.108/
g_3\right)\hfill \mbox{for} & \chi=\ax. \hfill \cr}
%\end{array}
\right. \label{sig1} \eeq
Here, $g_\alpha$ and $M_\alpha$ (with $\alpha=1,2,3$) are the
gauge coupling constants and gaugino masses respectively,
associated with the gauge groups $U(1)_{\rm Y}$, $SU(2)_{\rm L}$
and $SU(3)_{\rm C}$, $(k_\alpha)=(1.634,1.312,1.271)$ and
$(c_\alpha)=(33/5,27,72)$. Throughout our analysis we impose
universal initial conditions for the gaugino masses,
$M_\alpha(M_{\rm GUT})=M_{1/2}$ and gauge coupling constant
unification, i.e., $g_\alpha(M_{\rm GUT})=g_{\rm GUT}$.
\Eref{sig1} gives meaningful results only for $T>\Tc=10^4~\GeV$.
Towards lower values of $T$, non-relativistic ($T\ll m_i$)
contributions start playing an important role. In the case of
$\ax$, where $\Omega_{\ax} h^2$ takes cosmologically interesting
values for $T\ll m_i$, $C_{\ax}$ has been calculated numerically
in Ref.~\cite{GLPR}. In the latter case, $\Gm{i}$ with
$i=\gl,~\sq$ and $\tilde B$ are taken into account, also, using
the formulas of Refs.~\cite{axino, small}.

The impact of the KD phase on the relic density $\Omx= 2.748
\times 10^8\ Y^0_\chi\ m_{\chi}/\GeV$ can be found by solving
 Eqs.~(\ref{qeq}) and (\ref{nx}) numerically \cite{GLPR}.
However, we can get a clear picture of the results via
semi-analytical estimates \cite{GLPR}. In the high $T$ regime
($\Ti\gg \Tc$ and $\Tkr\gg \Tc$) we find, within a $10\%$
accuracy, that:
\beq \label{YhT}
 Y_\chi^0={n_{\chi}\over s}\simeq\left\{\matrix{ y^{\rm HT}_\sigma
\Ti\hfill & \mbox{in the SC}, \hfill \cr y^{\rm HT}_\sigma
\sqrt{{g^{\rm KR}_*/ g^{\rm I}_*}}\Tkr\ln\left(T_{\rm I}/T_{\rm
KR}\right)+ y^{\rm HT}_\sigma T_{\rm KR} \hfill &\mbox{in the
QKS,} \hfill \cr}\right. \eeq
where $s$ is the entropy density and $y_\sigma^{\rm
HT}=\sqrt{8g_{*}/45}\pi m_{\rm P }Y^{\rm eq2}C^{\rm HT}_\chi$ with
$Y^{\rm eq}={n^{\rm eq}/s}$. From these expressions we can easily
deduce that in QKS, $Y_\chi^0$ decreases with respect to its value
in the SC (being proportional to $\Ti$ in SC and to $\Tkr$ --
lower than $\Ti$ -- in QKS).

In the low $T$ regime ($\Ti\ll \Tc$ or $\Tkr\ll\Tc$), we find
cosmologically interesting solutions only in the case of $\ax$.
Focusing on the most intriguing possibility, $\Ti\gg\Ts=1~\TeV$
but $\Tkr\ll\Ts$, and for the benchmark values of $m_i$ used in
our analysis:
\beq \label{mi}
m_{\sq}=1~\TeV,~m_{\gl}=1.5~\TeV~~\mbox{and}~~m_{\tilde
B}=0.3~\TeV.\eeq
we can write simple empirical relations which reproduce rather
accurately the numerical results. In particular, in the SC, using
fitting technics, we get a relation with a $15\%$ accuracy,
namely: $\Omax=A\, m_{\ax}\,\left(1 +C\,\Ti
\right)\,e^{-B/\Ti}/f^2_a$ with $A=1.44\times
10^{24}~\GeV$,~$B=745.472~\GeV$ and $C=0.001/\GeV$. We observe
that $\Omax$ decreases sharply as $\Ti$ decreases due to the
exponential factor. In the QKS this suppression is avoided since
$\Omax=D\, {m_{\ax}/f^2_a\sqrt{\Omqns}}$ with
$D=9.26\times10^{17}~\GeV$. This relation reproduces the numerical
results with excellent accuracy. We observe that
$\Omax\propto1/\sqrt{\Omqns}$ or $\Omax\propto\Tkr$.

\begin{figure}[t]\centering\hspace*{0.2cm}
\includegraphics[width=6.cm,angle=-90]{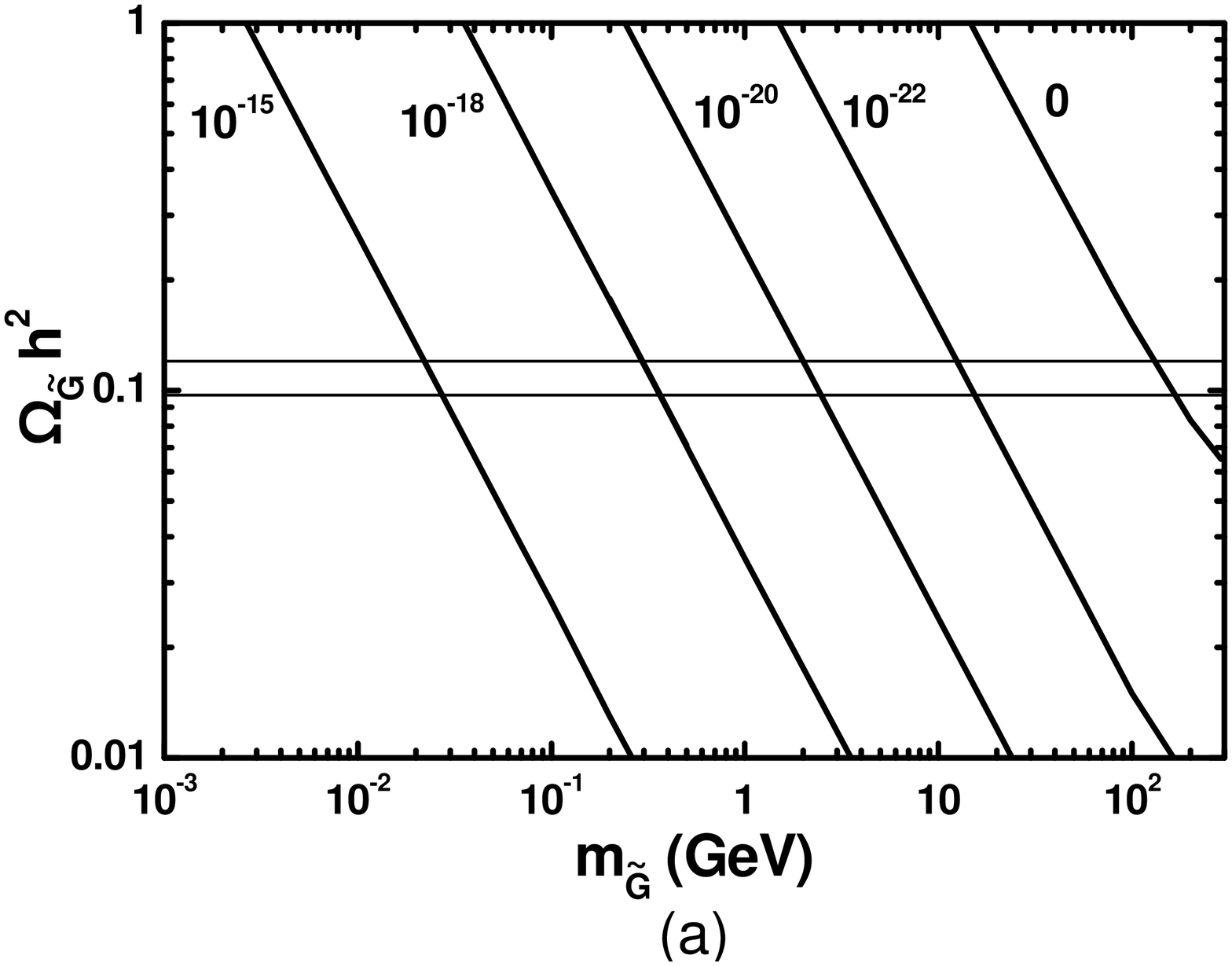}\hspace*{-1cm}
\includegraphics[width=6.cm,angle=-90]{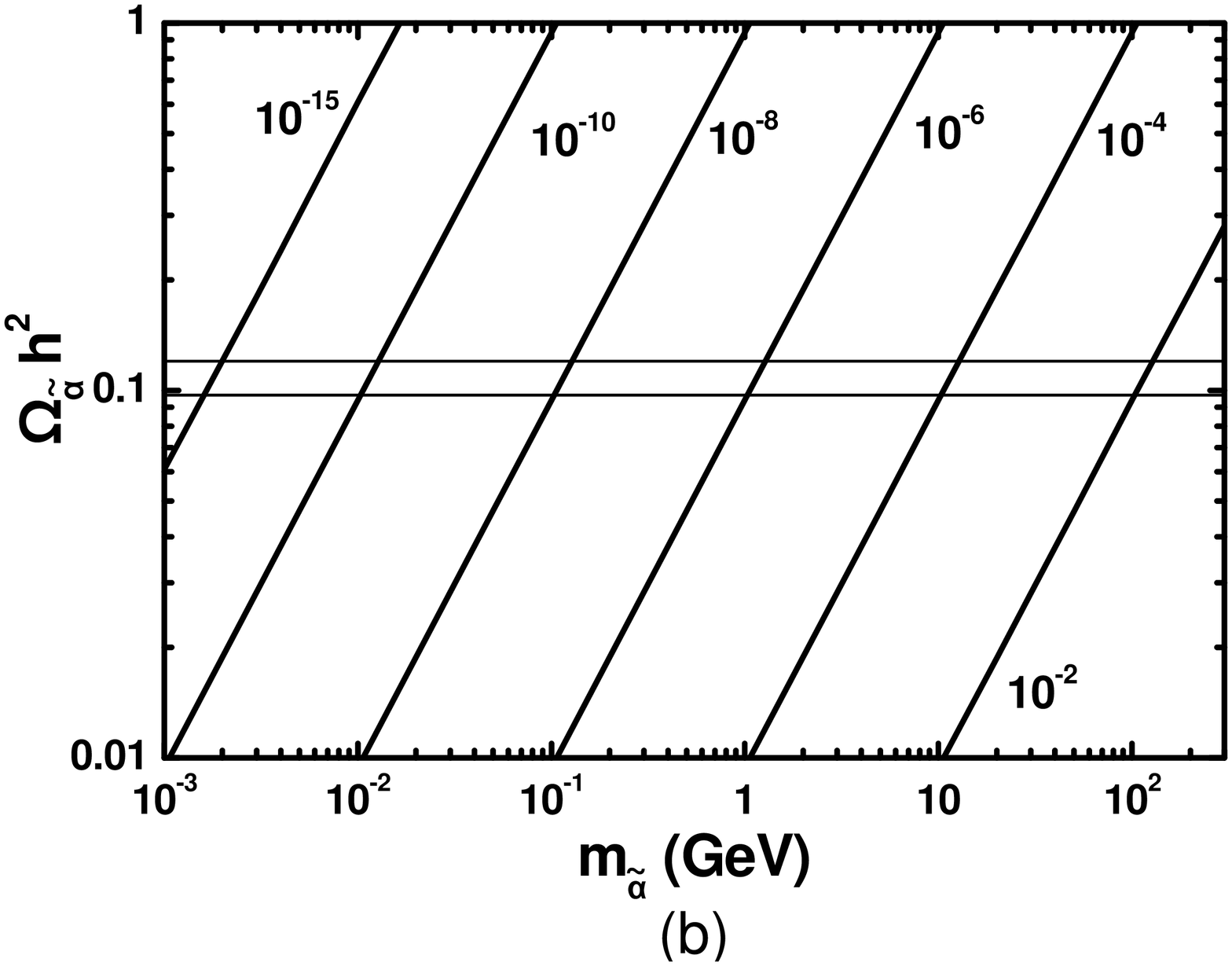}
\caption{\sl $\Omega_{\chi}h^2$ as a function of $m_\chi$
($\chi=\Gr$ [$\chi=\ax$]) for various $\Omega_q^{\rm NS}$'s,
indicated on the curves, $\Ti=10^9~\GeV$ and $M_{1/2}=0.7~\TeV$
[$f_a=10^{11}~\GeV$] (${\sf a}$ [${\sf b}$]). For $\Omega_q^{\rm
NS}>10^{-15}$, we take in our computation the values of $m_i$
indicated in Eq.~(\ref{mi}). The CDM bounds are also, depicted by
the two thin lines.} \label{om}
\end{figure}

In Figs.~\ref{om}-{$\sf (a)$} [\ref{om}-{$\sf (b)$}], we display
$\Omgr$ [$\Omax$] versus $m_{\Gr}$ [$m_{\ax}$] for
$M_{1/2}=0.7~\TeV$ [$f_a=10^{11}~\GeV$], $\Ti=10^9~\GeV$ and
various $\Omqns$'s indicated in the curves. We observe that $\Omx$
decreases as $\Omqns$ increases, $\Omgr\propto 1/m_{\Gr}$ and
$\Omax\propto m_{\ax}$.

%%%%%%%%%%%%%%%%%%%%%%%%%%%%%%%%%%%%%%%%%%%%%%%%%%%%%%%%%%%%%%%%%%%%
%%%%%%%%%%%%%%%%%%%%%%%%%%%%%%%%%%%%%%%%%%%%%%%%%%%%%%%%%%%%%%%%%%%%%%

\section{Quintessential Kination and $\Gr$-Constraint}

%%%%%%%%%%%%%%%%%%%%%%%%%%%%%%%%%%%%%%%%%%%%%%%%%%%%%%%%%%%%%%%%%%%%
\begin{figure}[!t]\centering\hspace*{0.2cm}
\includegraphics[width=6.cm,angle=-90]{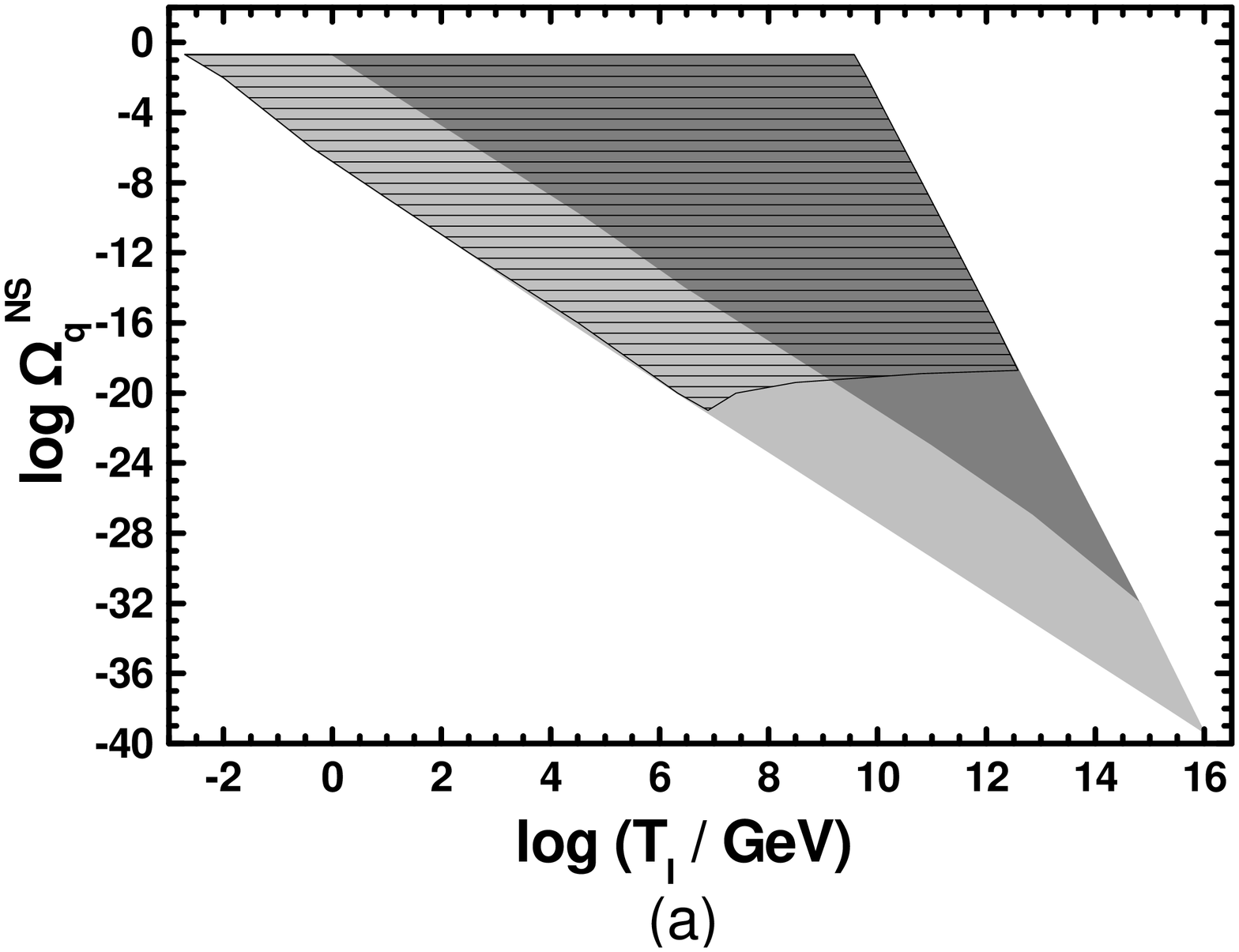}\hspace*{-1cm}
\includegraphics[width=6.cm,angle=-90]{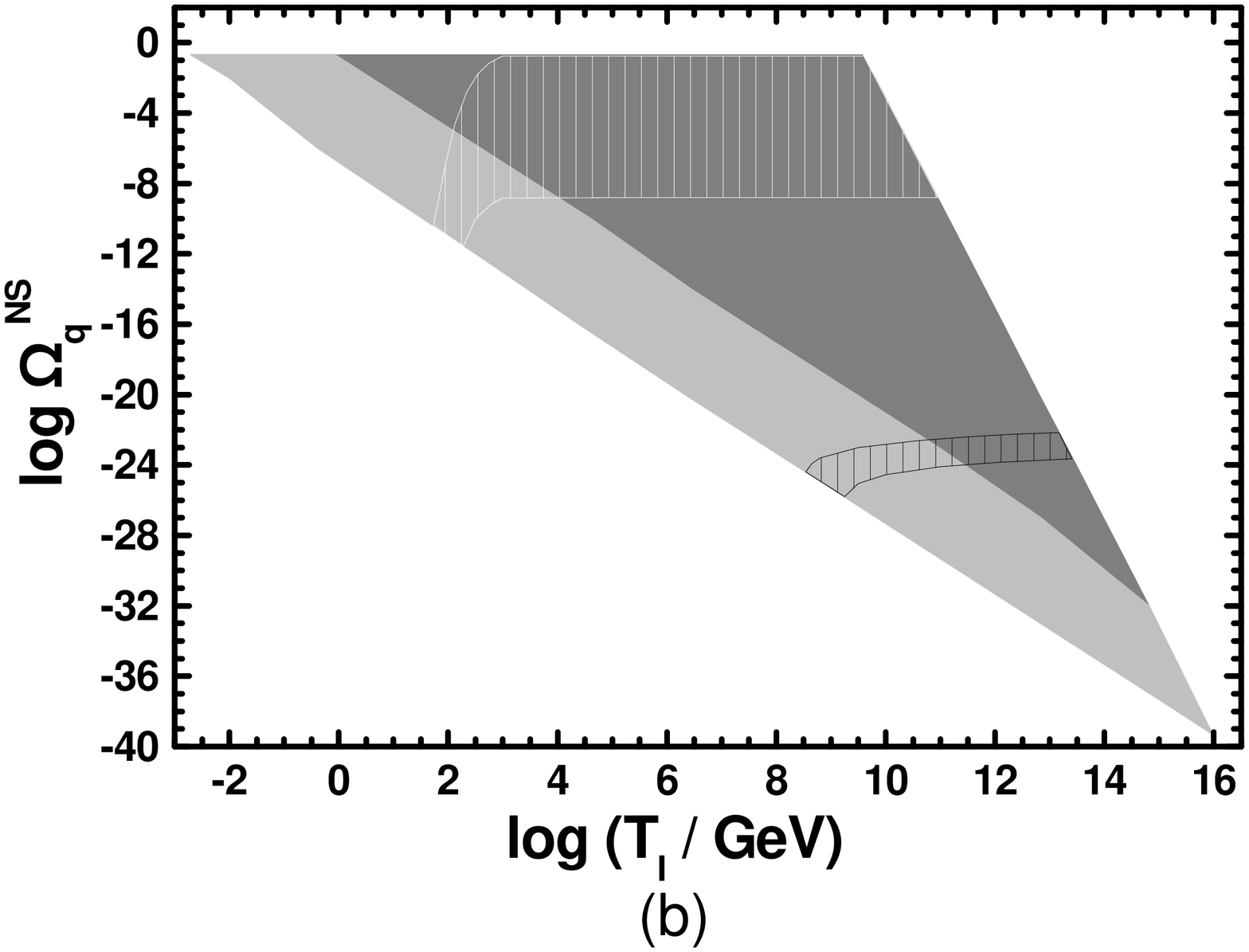}
\caption{\sl In the $\log\Ti-\log\Omqns$ plane we show the (gray
and lightly gray shaded) area, allowed by the quintessential
requirements \ref{domk} -- \ref{para}, and (a) the (black lined)
area allowed by the $\Gr$-constraint for $m_{\tilde
G}=M_{1/2}=0.5~\TeV$; (b) the (black lined) area, allowed by
\Eref{cdmb}, for $\Gr$-CDM with $m_{\Gr}=0.1~\TeV$ and $0.5\leq
M_{1/2}/\TeV\leq1$ or (white lined) area, allowed by \Eref{cdmb},
for $\ax$-CDM with $m_{\ax}=5~\GeV$, $m_i$'s of \Eref{mi} and
$10^{10}\leq f_a/\GeV\leq10^{12}$.}
%%%%%%%%%%%%%%%%
\label{OmTcdm}
\end{figure}

Unstable $\Gr$ can decay after the onset of NS, affecting by
unaccepted amounts the primordial abundances of light elements. To
avoid this, an upper bound on $Y_{\tilde G}(\Tns)$ can be
extracted as a function of $m_{\Gr}$ assuming that the hadronic
branching ratio of $\Gr$ is tiny. E.g., for $m_{\tilde
G}\simeq0.6~\TeV$, we obtain \cite{kohri} $Y_{\tilde
G}(\Tns)\lesssim10^{-14}$ which means that $T_{\rm I}\lesssim
4\times10^{7}~{\rm GeV}$ in the SC. This very restrictive upper
bound on $T_{\rm I}$ can be avoided in the QKS, where we can set
$T_{\rm I}=10^9~\GeV$. The upper bound on $Y_{\tilde G}(\Tns)$ can
be satisfied for $\Omega_q^{\rm NS}\gtrsim 10^{-21}$ or
$\Tkr\lesssim6.8\times10^{6}~{\rm GeV}$. The importance of a KD
era in evading the $\Gr$-constraint can be induced, also, by
\sFref{OmTcdm}{a}, where we show the allowed (black lined) area by
the $\Gr$-constraint in the $\log\Ti-\log\Omqns$ plane, for
$m_{\tilde G}=M_{1/2}=500~{\rm GeV}$. The (gray and lightly gray
shaded) area allowed by the quintessential requirements \ref{domk}
-- \ref{para} is also shown. We clearly see that the
$\Gr$-constraint can be totally eluded in the QKS even with tiny
values of $\Omega_q^{\rm NS}$ almost independently on $\Ti$.

\section{Quintessential Kination and $\Gr$ or $\ax$ CDM}

Stable $\chi$'s constitute good CDM candidates provided that their
relic density $\Omega_{\chi}h^2$ satisfies the CDM constraint
\cite{wmap}
\beq\label{cdmb} 0.097\lesssim \Omega_{\chi}h^2 \lesssim 0.12.\eeq
This constraint can be satisfied by both the $\Gr$ and $\ax$
thermal abundance. However, in the case of $\Gr$, $\Omqns$ is to
be tuned to extremely low values whereas in the case of $\ax$,
$\Omqns$ may be even close to its upper bound posed by the
condition \ref{nuc}. Indeed, in \sFref{OmTcdm}{b}, we present the
region in the $\log T-\log\Omqns$ plane allowed by both the
quintessential requirements, \ref{domk} -- \ref{para}, (gray and
lightly gray shaded area) and \Eref{cdmb} for $\Gr$-CDM (black
lined region) with $m_{\Gr}=100~\GeV$ and $0.5\leq
M_{1/2}/\TeV\leq1$ or $\ax$-CDM (white lined region) with
$m_{\ax}=5~\GeV$ and $10^{10}\leq f_{a}/\GeV\leq10^{12}$.
Obviously \Eref{cdmb} is met for $\ax$-CDM with much more natural
$\Omqns$'s than those required for $\Gr$-CDM. Therefore, $\ax$ is
more natural CDM candidate than $\Gr$ in the QKS.

\section{Conclusions}

We examined the impact of a KD epoch, generated by an
quintessential exponential model, to the thermal abundance of
$\Gr$ and $\tilde a$. The parameters of the quintessential model
($\lambda, \Ti, \Omqns$) were confined so as
$0.5\leq\Omega_q(\Ti)\leq1$ and were constrained by using current
observational data originating from NS, the acceleration of the
universe, the inflationary scale and the DE density parameter. We
found that $0<\lambda<0.9$ and studied the allowed region in the
($\Ti, \Omqns$)-plane. For unstable $\tilde{G}$, the
$\tilde{G}$-constraint poses a lower bound on $\Omqns$ (almost
independent of $\Ti$). The CDM constraint can be satisfied by the
$\Gr$ thermal abundance for extremely low $\Omqns$'s. On the
contrary, this constraint can be fulfilled by the $\ax$ thermal
abundance with much larger $\Omqns$'s, making $\ax$ a very good
CDM candidate.

\begin{theacknowledgments}
The research of S.L is funded by the FP6  Marie Curie Excellence
Grant MEXT-CT-2004-014297. The work of M.E.G, C.P and J.R.Q is
supported by the Spanish MEC projet FPA2006-13825 and the projet
P07FQM02962 funded by ``Junta de Andalucia''.
\end{theacknowledgments}

\newcommand{\arxiv}[1]{{\tt arXiv:#1}}

\newcommand{\hepph}[1]{{\tt hep-ph/#1}}
\newcommand{\hepex}[1]{{\tt hep-ex/#1}}
\newcommand{\astroph}[1]{{\tt astro-ph/#1}}
\newcommand{\hepth}[1]{{\tt hep-th/#1}}
\newcommand{\grqc}[1]{{\tt gr-qc/#1}}
\newcommand{\etal}{{\it et al.\/}}

\newcommand\astp[3]{\emph{ Astropart.\ Phys.\ }{\bf #1}, #3 (#2)}

\newcommand\apj[3]
        {\emph{ Astropart.\ J. \ }{\bf #1}, #3 (#2)}

\newcommand\jhep[3]
        {\emph{ J. High Energy Phys.\ }{\bf #1}, #3 (#2)}

\newcommand\jcap[3]
        {\emph{ J. Cosmol. Astropart. Phys.\ }{\bf #1}, #3 (#2)}

\newcommand\npb[3]
        {\emph{ Nucl.\ Phys.\ }{\bf B#1}, #3 (#2)}

\newcommand\plb[3]
        {\emph{Phys.\ Lett.\ }{B \bf #1}, #3 (#2)}

\newcommand\prd[3]
        {\emph{ Phys.\ Rev.\ }{D \bf #1}, #3 (#2)}

\newcommand\prep[3]
        {\emph{ Phys.\ Rep.\ }{\bf #1}, #3 (#2)}
\newcommand\prl[3]
        {\emph{ Phys.\ Rev.\ Lett.\ }{\bf #1}, #3 (#2)}

%%%%%%%%%%%%%%%%%%%%%% HEP's%%%%%%%%%%%%%%%%%%%%%%%%%%%%%%%%%%%%%%%%%%%%

\end{document}